\documentclass[12pt, a4paper, superscriptaddress, final]{iopart}

\expandafter\let\csname equation*\endcsname\relax
\expandafter\let\csname endequation*\endcsname\relax

\usepackage{amsmath,amssymb}
\usepackage{graphicx}
\usepackage{dcolumn}
\usepackage{bm}
\usepackage{hyperref}
\usepackage{url}
\usepackage{braket}
\usepackage{color}

\usepackage{enumitem}

\usepackage{titlecaps}
\Addlcwords{the of and to}

\usepackage{longtable}
\usepackage{makecell}
\usepackage{etoolbox}

\makeatletter
\newcommand{\mainmatter}{%
  \setcounter{footnote}{0}%
  \patchcmd{\@makefntext}{\fnsymbol}{\arabic}{}{}%
  \patchcmd{\@thefnmark}{\fnsymbol}{\arabic}{}{}%
  \def\@makefnmark{\textsuperscript{\arabic{footnote}}}%
}
\makeatother

\begin{document}


\title{Photon centroids and their subluminal propagation}

\author{Konstantin Y. Bliokh}
\address{Theoretical Quantum Physics Laboratory, Cluster for Pioneering Research, RIKEN, Wako-shi, Saitama 351-0198, Japan}
\address{Centre of Excellence ENSEMBLE3 Sp. z o.o., 01-919 Warsaw, Poland}
\address{Donostia International Physics Center (DIPC), Donostia-San Sebasti\'{a}n 20018, Spain}

\begin{abstract}
We examine properties and propagation of the energy-density and photon-probability centroids of electromagnetic wavepackets in free space. In the second-order paraxial approximation, both of these centroids propagate with the same subluminal velocity because of the transverse confinement of the wavepacket and its diffraction. The tiny difference between the energy and probability centroid velocities appears only in the fourth order. We consider three types of wavepackets: Gaussian, Bessel, and non-diffracting Bessel. In all these cases, the subluminal propagation is clearly visible in the intensity distributions and can be measured experimentally in both classical-light and single-photon regimes. For Gaussian wavepackets, the half-wavelength delay is accumulated after propagation over about 12 Rayleigh lengths.    
\end{abstract}

%
%
%
%
%



\section{Introduction}

Subluminal and superluminal velocities in the propagation of light has intrigued scientists for several decades \cite{Brillouin_book,Chiao1997,Winful2006,Boyd2002,Khurgin2010, Berry2010, Berry2012, Bliokh2013NJP_II, Asano2016, Giovannini2015}. There are different velocities characterizing propagation of light (phase, group, signal, energy-transfer, etc.) and different mechanisms for the subluminal and superluminal propagations: interactions with matter \cite{Chiao1997, Winful2006, Boyd2002, Khurgin2010, Asano2016}, local phase gradients in structured free-space light \cite{Berry2010, Berry2012, Bliokh2013NJP_II}, tilt of the plane-wave Fourier components in a transversely-confined light beams \cite{Giovannini2015}, and specially space-time structured wavepackets \cite{Yessenov2022}. 

In this work, we examine propagation of the energy and photon-probability centroids of regular optical wavepackets and analyse their subluminal velocities. 
This phenomenon of subluminal propagation originates from the transverse confinement of the wavepackets, and hence is related to the mechanism considered in Ref.~\cite{Giovannini2015}. However, while Ref.~\cite{Giovannini2015} considered single-photon wavepackets within monochromatic (i.e., longitudinally delocalized) optical-beam setup, we consider polychromatic transversely and longitudinally confined wavepackets with well defined centroids. Thus, our problem is equally well defined in the single-photon and classical-light frameworks.   

A fundamental difficulty with the definition of a photon centroid is that the photon wavefunction, and, hence, the probability density, are ill-defined in real space \cite{CT, BB1996}. (This is a consequence of the Weinberg-Witten theorem for massless spin-1 particles \cite{Weinberg1980}.) Only the {\it energy} density is well defined, and the corresponding {\it energy centroid} is calculated straightforwardly. It moves rectilinearly and uniformly in free space, as a consequence of the conservation of the {\it boost momentum}, i.e., spatiotemporal components of the relativistic angular momentum tensor \cite{LLfield,Bliokh2013NJP}. 
Nonetheless, despite the ill-defined probability density, the {\it probability centroid} can be determined as the expectation value of the position operator in the momentum representation \cite{Smirnova2018, Bliokh2023}. 

In this work, we explicitly calculate the motion of both the energy and probability centroids of a `semiclassical' (i.e., large as compared with the central wavelength) electromagnetic wavepacket in free space. We show that both centroids move rectilinearly and uniformly with velocities slightly less than the speed of light $c$. The difference between these velocities and $c$ is of the second order in the small paraxiality parameter (the characteristic polar angle of the wavevectors in the wavepacket), while the difference between the energy and probability centroid velocities is of the fourth order. We consider different examples of wavepackets: Gaussian, spatiotemporal Bessel, and non-diffracting spatiotemporal Bessel.

\section{Energy and photon-probability centroids}

The first-quantization wave formalism for photons is the classical Maxwell electromagnetism \cite{Laporte1931, Oppenheimer1931, BB1996}. The energy centroid of an electromagnetic wavepacket is
\begin{equation}
\label{eq1}
{\bf R}_E = \frac{\int W{\bf r}\, d^3{\bf r}}{\int W d^3{\bf r} } \,,
\end{equation}
where $W = \left({\bf E}^2+{\bf H}^2\right)/2$ is the energy density, whereas ${\bf E}({\bf r},t)$ and ${\bf H}({\bf r},t)$ are the real-valued electric and magnetic fields, and we use the Gaussian units omitting inessential numerical constants. The velocity of the energy centroid propagation is obtained by differentiating (\ref{eq1}) with respect to time:
\begin{equation}
\label{eq2}
{\bf V}_E =
\frac{\partial}{\partial t}\frac{\int W{\bf r}\, d^3{\bf r}}{\int W d^3{\bf r} } = - \frac{\int ({\bm \nabla} \cdot {\bm \Pi})\, {\bf r}\, d^3{\bf r}}{\int W d^3{\bf r} } = \frac{\int {\bf \Pi}\, d^3{\bf r}}{\int W d^3{\bf r} } = {\bf const} \,.
\end{equation}
Here ${\bf \Pi} = c({\bf E}\times{\bf H}) = c^2{\bf P}$ is the energy flux (Poynting vector), ${\bf P}$ is the   kinetic momentum density, and we used the energy and momentum conservation laws \cite{LLfield, Jackson}:  
\begin{equation}
\label{eq3}
\frac{\partial W}{\partial t} + {\bm \nabla} \cdot {\bm \Pi} = 0 \,, \quad
\frac{\partial}{\partial t} \int W d^3{\bf r} = 0\,, \quad 
\frac{\partial}{\partial t} \int {\bm \Pi} \,d^3{\bf r} = {\bf 0}\,.
\end{equation}

Equation~(\ref{eq2}) shows that the energy centroid moves uniformly and rectilinearly in free space. 
The energy-centroid velocity ${\bf V}_E$ is given by the ratio of the integral Poynting vector (i.e., momentum multiplied by $c^2$) to the integral energy. This form is equivalent to the velocity of a free relativistic classical particle: ${\bf v} = c^2 {\bf p}/{\cal E}$ \cite{LLfield}. In some electromagnetic problems (e.g., for quasi-1D transversely-localized monochromatic guided modes), the ratio of the integral Poynting vector to the integral energy is associated with the group velocity ${\bf v}_g = \partial \omega ({\bf k}) / \partial {\bf k}$ \cite{Snyder, Nkoma1974, Bliokh2017PRL, Picardi2018}. However, as we will see, the situation becomes more sophisticated for transversely and longitudinally localized free-space wavepackets. Note also that Eq.~(\ref{eq2}) follows from the conservation of the integral electromagnetic boost momentum \cite{LLfield, Bliokh2013NJP, Smirnova2018}:  
\begin{equation}
\label{eq4}
 \int ({\bf r} W - t {\bm \Pi}) \,d^3{\bf r} = {\bf const}\,, 
 \end{equation}

Let us now define the photon probability centroid. Since the photon wavefunction is well defined only in the momentum representation, we use the complex Fourier (plane-wave) amplitudes of the electric and magnetic fields, $\tilde{\bf E}({\bf k})e^{-i\omega({\bf k})t}$ and $\tilde{\bf H}({\bf k})e^{-i\omega({\bf k})t}$, satisfying $\tilde{\bf H} = c\,{\bf k} \times \tilde{\bf E}/\omega$ and $ \omega ({\bf k}) = c k$. In all calculations below, the magnetic- and electric-field amplitudes make equal contributions, so that we can use only the electric-field amplitudes. Then, the photon wavefunction in the momentum representation is given by ${\bm \psi} ({\bf k},t) \propto \tilde{\bf E}({\bf k})e^{-i\omega({\bf k})t} / \sqrt{\omega({\bf k})}$ \cite{CT, BB1996, Smirnova2018, Bliokh2023}, and the probability centroid is given by the normalized expectation value of the canonical position operator $\hat{\bf r} = i {\bm \nabla}_{\bf k}$:
\begin{equation}
\label{eq5}
{\bf R} = \frac{\langle {\bm \psi} | \hat{\bf r} | {\bm \psi} \rangle}{\langle {\bm \psi} | {\bm \psi} \rangle} =
\frac{ \int  \omega^{-1} \tilde{\bf E}^*e^{i\omega t} \cdot (i{\nabla}_{\bf k}) \tilde{\bf E}e^{-i\omega t} \, d^3{\bf k}}{\int \omega^{-1} |\tilde{\bf E}|^2 \, d^3{\bf k}}\,.
 \end{equation}
This centroid has been previously employed for the analysis of the intrinsic and extrinsic contributions to the orbital angular momentum of vortex wave states \cite{ Bliokh2012PRL, Bliokh2012, Smirnova2018, Bliokh2023}. Calculating the time derivative of Eq.~(\ref{eq5}), we find the photon probability-centroid velocity:
\begin{equation}
\label{eq6}
{\bf V} = \frac{\partial}{\partial t} \frac{ \int  \omega^{-1} \tilde{\bf E}^*e^{i\omega t} \cdot (i{\nabla}_{\bf k}) \tilde{\bf E}e^{-i\omega t} \, d^3{\bf k}}{\int \omega^{-1} |\tilde{\bf E}|^2 \, d^3{\bf k}} =
\frac{ \int  \omega^{-1} |\tilde{\bf E}|^2 {\bf v}_g \, d^3{\bf k}}{\int \omega^{-1} |\tilde{\bf E}|^2 \, d^3{\bf k}} \equiv \langle {\bf v}_g \rangle= {\bf const}\,.
 \end{equation}
Here ${\bf v}_g = \partial \omega/\partial {\bf k} = c{\bf k}/k$ is the ${\bf k}$-dependent group velocity. Equation (\ref{eq6}) is manifestly time-independent, which indicates the uniform rectilinear propagation of the probability centroid with the mean group velocity $\langle {\bf v}_g \rangle$. 

In the photon-wavefunction ${\bf k}$-space formalism, the integral electromagnetic energy and momentum take the forms $\int W d^3{\bf r} \propto \langle {\bm \psi} | \omega | {\bm \psi} \rangle$ and $\int {\bf P} d^3{\bf r} \propto \langle {\bm \psi} | {\bf k} | {\bm \psi} \rangle$, so the energy centroid (\ref{eq1}) and its velocity (\ref{eq2}) can be written as 
\begin{equation}
\label{eq7}
{\bf R}_E = \frac{\langle {\bm \psi} | \omega \hat{\bf r} | {\bm \psi} \rangle}{\langle {\bm \psi} | \omega | {\bm \psi} \rangle} =
\frac{ \int  \tilde{\bf E}^*e^{i\omega t} \cdot (i{\nabla}_{\bf k}) \tilde{\bf E}e^{-i\omega t} \, d^3{\bf k}}{\int |\tilde{\bf E}|^2 \, d^3{\bf k}}\,,
 \end{equation}
\begin{equation}
\label{eq8}
{\bf V}_E = \frac{ \int  |\tilde{\bf E}|^2 {\bf v}_g \, d^3{\bf k}}{\int  |\tilde{\bf E}|^2 \, d^3{\bf k}}
\equiv \langle {\bf v}_g \rangle_E = {\bf const}\,.
 \end{equation}
Here, the averaged group velocity $\equiv \langle {\bf v}_g \rangle_E$ is weigthed by the electric-field rather than wavefunction intensity. The velocities of the energy and probability centroids, in general, differ from each other due to the extra $\omega ({\bf k})$ factors in the integrals (\ref{eq5})--(\ref{eq8}).

\section{Propagation of the centroids of paraxial wavepackets}

\subsection{Gaussian wavepacket in the second-order approximation}

\begin{figure}[!t]
\includegraphics[width=\linewidth]{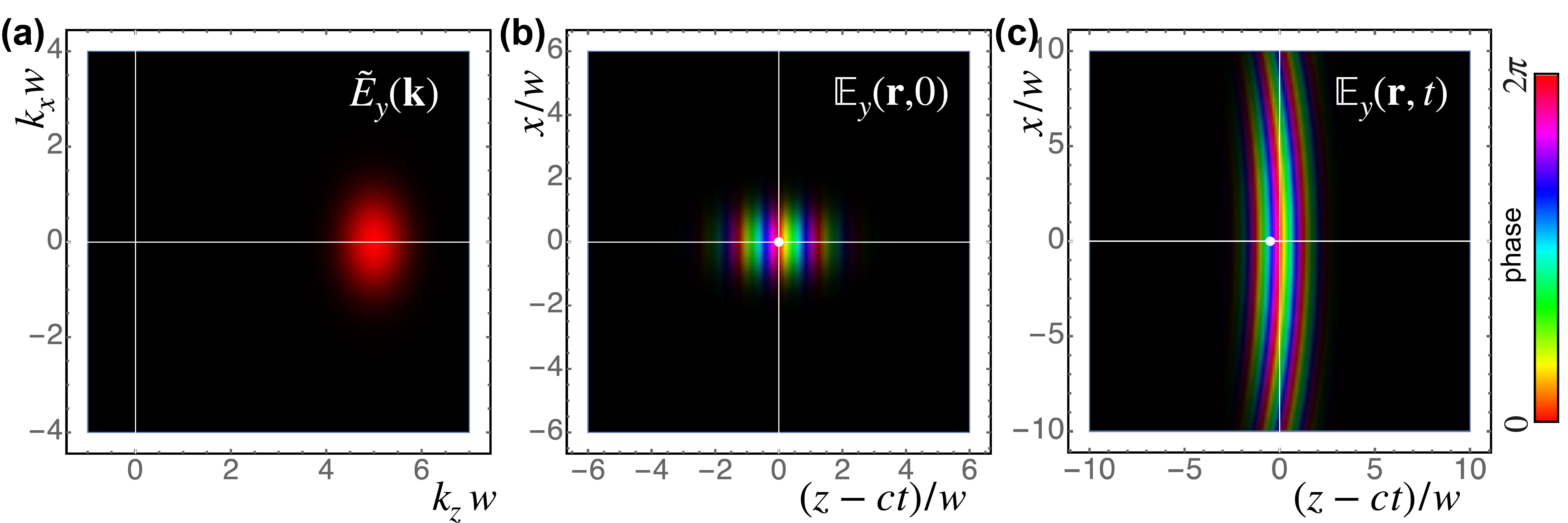}
\caption{Plane-wave Fourier spectrum (\ref{eq9}) and real-space field (\ref{eq12}) of the Gaussian wavepacket with $k_0 w = 5$ and $k_0 l = 7.5$. Here and hereafter, we use representation of complex fields with brightness proportional to its intensity and colours indicating the phase \cite{Thaller_book}. The panels (b) and (c) show the real-space field at $t=0$ and after propagation over 10 Rayleigh lengths: $ct = 10 k_0 w^2$. The white dot indicates the position of the energy and probability centroids $Z\simeq Z_E < ct$ propagating with  subluminal velocity (\ref{eq11}).}
\label{fig1}
\end{figure}

We now examine propagation of paraxial electromagnetic wavepackets in free space. First, consider a Gaussian wavepacket, which propagates along the $z$-axis and is localized in the $(x,z)$-plane but, for the sake of simplicity, infinite and uniform in the $y$-direction. This enables us to omit inessential $y$-integrals and assume uniform $y$-polarization with a single electric-field component $E_y$ satisfying the transversality condition ${\bm \nabla} \cdot {\bf E} = \partial E_y / \partial y =0$. 
We write the Gaussian Fourier spectrum, Fig.~\ref{fig1}(a), as
\begin{equation}
\label{eq9}
\tilde{E}_y ({\bf k}) \propto \exp\!\left[ -\frac{(k_z-k_0)^2 l^2}{2} - \frac{k_x^2 w^2}{2} \right],
 \end{equation}
where $k_0$ determines the central wavevector 
${\bf k}_0 = k_0 \bar{\bf z}$ (the overbar denotes the unit vector of the corresponding axis), whereas $l \gg k_0^{-1}$ and $w \gg k_0^{-1}$ denote the $z$-length and $x$-width of the `semiclassical' wavepacket.

In the second-order approximation in small parameters $\kappa_z = (k_z - k_0)/k_0$ and $\kappa_x = k_x / k_0$, using relations $\omega = c k$ and ${\bf v}_g = c {\bf k}/k$, we find:
\begin{align}
\label{eq10}
& \omega^{-1} \simeq \omega_0^{-1}\! \left( 1 - \kappa_z + \kappa_z^2 - \frac{\kappa_x^2}{2} \right), \quad
{\bf v}_g \simeq c \!\left( 1 - \frac{\kappa_x^2}{2} \right)\! \bar{\bf z} + c \left(\kappa_x - \kappa_x \kappa_z \right) \bar{\bf x}\,, \nonumber \\
& \omega^{-1} {\bf v}_g \simeq \omega_0^{-1} c \left( 1 - \kappa_z + \kappa_z^2 - {\kappa_x^2} \right) \bar{\bf z} + \omega_0^{-1} c \left(\kappa_x - 2 \kappa_x \kappa_z \right) \bar{\bf x}\,.
 \end{align}
Substituting expressions (\ref{eq9}) and (\ref{eq10}) into Eqs.~(\ref{eq6}) and (\ref{eq8}) and calculating standard integrals over $(\kappa_x, \kappa_z)$, we derive:
\begin{equation}
\label{eq11}
{\bf V} = {\bf V}_E \simeq c \left( 1 - \frac{1}{4 k_0^2 w^2} \right)\! \bar{\bf z}\,. 
\end{equation}

Thus, in the second-order approximation, the energy and probability centroids propagate with the same velocity which is slightly less than speed of light. This effect originates from the tilt of the wavevectors in the transversely-localized wavepacket. That is why it is determined by the transversal width $w$ and can be equally measured on a single-photon level in $z$-delocalized wave beams \cite{Giovannini2015}.
To estimate the magnitude of this effect, note that propagation over 12 Rayleigh lengths $z_R = k_0 w^2$, $t = 12 k_0 w^2 / c$, yields the centroid delay of about half the central wavelength: $\Delta Z = (V-c)t \simeq - 3/ k_0 \simeq - \lambda_0/2$. This is quite feasible for experimental measurements.

Figure~\ref{fig1} shows the Fourier spectrum $\tilde{E}_y ({\bf k})$ of the Gaussian beam, Eq.~(\ref{eq9}), as well as its real-space intensity-phase distribution via complex field \cite{Siegman_book}
\begin{equation}
\label{eq12}
{\mathbb E}_y ({\bf r},t) \propto \exp\!\left[ -\frac{(z-ct)^2}{2 l^2} - \frac{x^2 }{2(w^2+i c t/k_0)} + i k_0 (z-ct)\right],
 \end{equation}
This field is the complex Fourier transform of $\tilde{E}_y ({\bf k})$, while the real-valued field is $E_y = {\rm Re}\, {\mathbb E}_y$. At $z=t=0$, the field presents the usual Gauassian wavepacket, Fig.~\ref{fig1}(b), but after propagation over several Rayleigh lengths diffraction transforms it into an arc-like far-field shape, Fig.~\ref{fig1}(c). The edges of this shape propagate slower than its centre at $x=0$, so that the resulting centroid becomes shifted to $Z - ct < 0$. This corresponds to the subluminal propagation with the velocity (\ref{eq11}). 

\subsection{Bessel wavepacket in the fourth-order approximation}

\begin{figure}[!t]
\includegraphics[width=\linewidth]{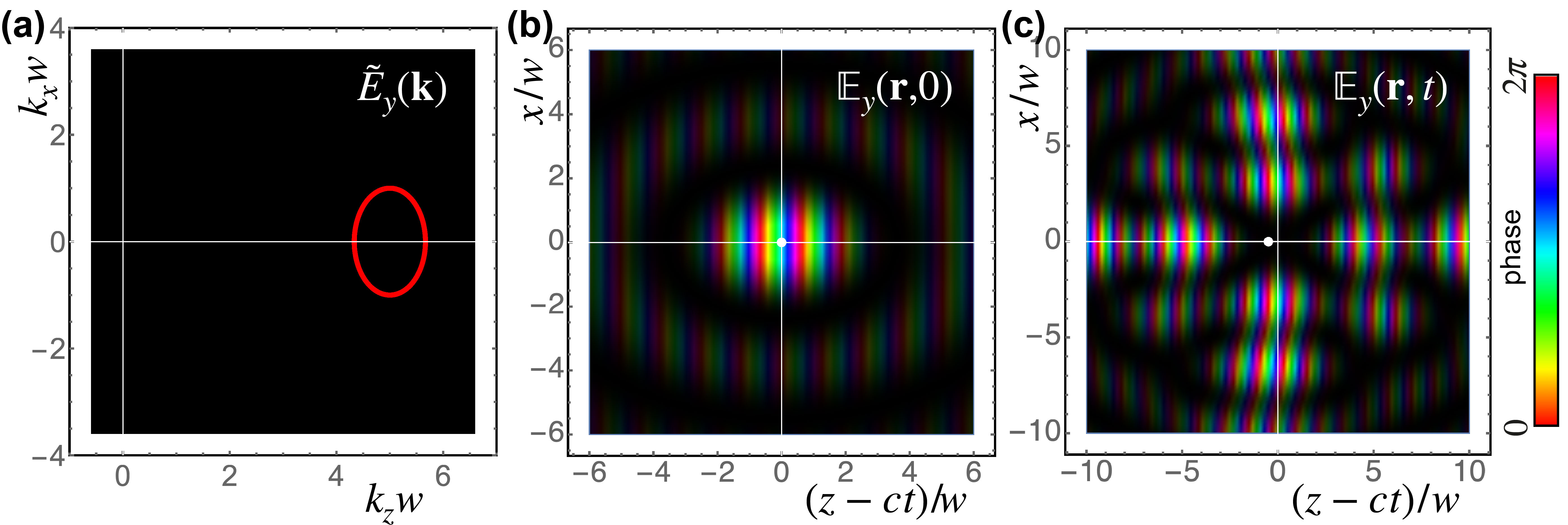}
\caption{Same as in Fig.~\ref{fig1} but for the Bessel-type wavepacket (\ref{eq13}) and (\ref{eq18}) with $m=0$ and the same values of parameters $w$, $l$, and $t$.}
\label{fig2}
\end{figure}

Let us examine another type of a wavepacket and perform calculation in a higher-order approximation. Bessel-type wavepackets provide the simplest example for ${\bf k}$-space calculations. Their spectrum can be presented by an ellipse centered at ${\bf k}_0 = k_0 \bar{\bf z}$ in the $(k_x,k_z)$ plane \cite{Bliokh2023,Bliokh2021}, Fig.~\ref{fig2}(a):
\begin{equation}
\label{eq13}
k_z = k_0 + l^{-1} \cos \phi \,, ~~
k_x = w^{-1} \sin \phi \,, ~~
E_y \propto \exp(im\phi)\,,
\end{equation}
where $\phi$ is a parameter such that $\varphi = \arctan\!\left(\dfrac{l}{w}\tan\phi\right)$ is the azimuthal angle with respect to the ellipse center. Here we consider Bessel wavepackets with the spatiotemporal vortex in the $(x,z)$ [or $(x,t)$] plane characterized by the integer quantum number $m = 0, \pm1, \pm 2, ...$. The zeroth-order wavepacket with $m =0$ does not have a vortex. 
Because of the one-dimensional character of the Fourier spectrum, the ${\bf k}$-space integration should be performed over the single azimuthal coordinate $\phi$.

Similar to the Gaussian wavepacket case, we first approximate the main quantities with respect to small parameters $(\kappa_x,\kappa_z) = (k_0^{-1}w^{-1} \sin\phi, k_0^{-1}l^{-1} \cos\phi)$, but now in the fourth-order approximation. This results in the higher-order corrections to Eqs.~(\ref{eq10}):
\begin{align}
\label{eq14}
& \omega^{-1} \to  \omega^{-1}  + \omega_0^{-1}\! \left( - \kappa_z^3 + \frac{3}{2}\kappa_z \kappa_x^2 + \kappa_z^4 - 3\kappa_z^2\kappa_x^2 +\frac{3}{8}\kappa_x^4 \right), \nonumber \\
& {\bf v}_g \to  {\bf v}_g + c \!\left( \kappa_x^2 \kappa_z - \frac{3}{2}\kappa_x^2\kappa_z^2 +\frac{3}{8}\kappa_x^4 \right)\! \bar{\bf z} 
 + c\! \left(\kappa_z^2 \kappa_x -\frac{\kappa_x^3}{2}  -\kappa_x \kappa_z^3 +\frac{3}{2}k_x^3 k_z \right)\! \bar{\bf x}\,, \nonumber \\
& \omega^{-1} {\bf v}_g \to \omega^{-1} {\bf v}_g + \omega_0^{-1} c \left( - \kappa_z^3 + 3 \kappa_x^2 \kappa_z + \kappa_z^4 - 6 \kappa_x^2 \kappa_z^2 + \kappa_x^4
\right) \bar{\bf z} 
\nonumber \\
& \qquad + \omega_0^{-1} c \left( 3\kappa_x \kappa_z^2 - \kappa_x^3 - 4 \kappa_x \kappa_z^3 + 4\kappa_x^3 \kappa_z \right) \bar{\bf x}\,.
 \end{align}
 Obviously, the $x$-components in these equations are antisymmetric functions of either $\kappa_x$ or $\kappa_z$ and all the integrals of these vanish. Substituting Eqs.~(\ref{eq13}) and (\ref{eq14}) into Eqs.~(\ref{eq6}) and (\ref{eq8}) and calculating the integrals of their $z$-components over $\phi$, we derive:
\begin{align}
\label{eq15}
{\bf V}_E \simeq c \left( 1 - \frac{1}{4 k_0^2 w^2} - \frac{3}{16 k_0^4 w^2l^2} + \frac{9}{64 k_0^4 w^4} \right)\! \bar{\bf z}\,, \\
\label{eq16}
{\bf V} \simeq c \left( 1 - \frac{1}{4 k_0^2 w^2} - \frac{1}{4 k_0^4 w^2l^2} + \frac{11}{64 k_0^4 w^4} \right)\! \bar{\bf z}\, ,
\end{align}

\begin{figure}[!t]
\includegraphics[width=\linewidth]{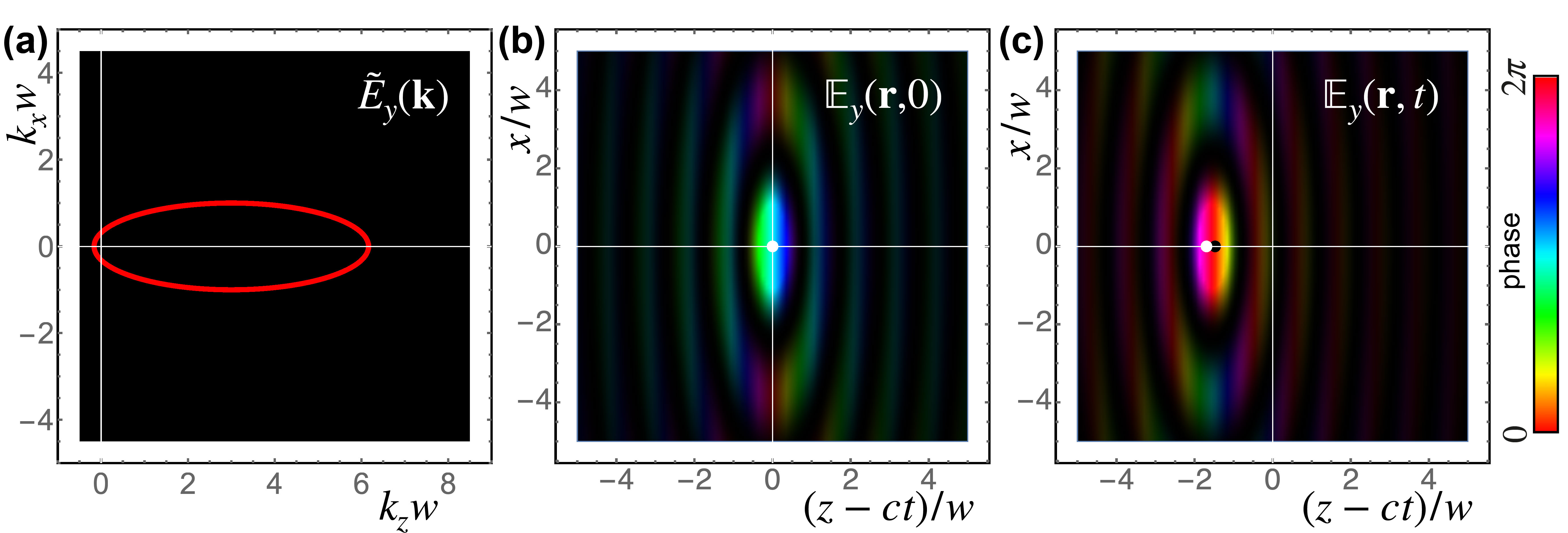}
\caption{Same as in Figs.~\ref{fig1} and \ref{fig2} but for the ultrashort (subwavelength) non-diffracting Bessel wavepacket (\ref{eq13}), (\ref{eq18}), and (\ref{eq19}). The parameters are: $m=0$, $k_0 w =3$, $k_0 l \simeq 0.95$, and $ct = 10 k_0 w^2$ in the panel (c). The white and black dots indicate positions of the probability and energy centroids $Z <ct$ and $Z_E <ct$, respectively, propagating with subluminal velocities (\ref{eq15})--(\ref{eq17}).}
\label{fig3}
\end{figure}

These equations show that the velocities of the energy and probability centroids differ in the fourth order in the semiclassical parameters:
\begin{align}
\label{eq17}
\delta {\bf V} \equiv {\bf V}_E - {\bf V} \simeq c \left(\frac{1}{16 k_0^4 w^2l^2} - \frac{1}{32 k_0^4 w^4} \right)\! \bar{\bf z}\,.
\end{align}
The transverse confinement of the wavepacket, $w$, is again crucial for this difference. As shown in Ref.~\cite{Bliokh2023}, the presence of a spatiotemporal vortex induces transverse $m$-dependent shifts of the energy and probability centroids, $X_E$ and $X$, but the above equations show that it does not affect the rectilinear $z$-propagation of these centroids. Therefore, in Figs.~\ref{fig2} and \ref{fig3} we show the simplest zeroth-order wavepckets with $m=0$.

Figure~\ref{fig2} shows the real-space evolution of the Bessel wavepacket numerically calculated via complex Fourier transform of the spectrum (\ref{eq13}):
\begin{equation}
\label{eq18}
{\mathbb E}_y ({\bf r},t) \propto \int_0^{2\pi} \tilde{E}_y (\phi) \exp\!\left[ i {\bf k}(\phi)\cdot {\bf r} - i \omega(\phi) t\right] d\phi\,.
 \end{equation}
At $z=t=0$ it has the elliptically deformed radial Bessel-function intensity distribution, Fig.~\ref{fig2}(b), but with propagation it undergoes strong deformations (due to the `temporal diffraction' \cite{Bliokh2021}) into a ring-like intensity distribution mimicking the elliptical Fourier spectrum, Fig.~\ref{fig2}(c). Akin to the Gaussian case, one can notice the acr-like deformations within this ring-like structure, which are responsible for the shift of the centroids to the subluminal area $z-ct<0$. In the second-order approximation the energy and probability centroids propagate with the same subluminal velocity (\ref{eq11}); the fourth-order difference (\ref{eq17}) is still unnoticeable in Fig.~\ref{fig2}.

In the above calculations we assumed a uniform distribution of the density of states over the azimuthal parameter $\phi$. For the uniform distribution over the azimuthal angle $\varphi$, all the integrals involve extra $d\varphi/d\phi$ factors. This modifies some numerical factors in Eqs.~(\ref{eq15})--(\ref{eq17}), but does not affect the qualitative picture.   

\subsection{Non-diffracting Bessel wavepacket}

A remarkable example is provided by the Bessel wavepacket described by the spectrum (\ref{eq13}) with 
\begin{equation}
\label{eq19}
l=\frac{w}{\sqrt{1+k_0^2 w^2}}\,,
 \end{equation}
shown in Fig.~\ref{fig3}(a). Such spectrum can be transformed to a circle centred at ${\bf k}={\bf 0}$ by a suitable Lorentz boost along the $z$-direction \cite{Bliokh2012,Bliokh2021}. This means that such Bessel wavepacket represents a Lorentz boost of a standing Bessel circular wave, and hence its intensity distribution {\it does not diffract} during the propagation \cite{Bliokh2012,Bliokh2021}, Fig.~\ref{fig3}(b,c), providing a linear 2D analogue of `light bullets' \cite{Silberberg1990,Minardi2010}. The price for this is that the elliptical Fourier spectrum of such non-diffracting Bessel wavepacket embraces the ${\bf k}={\bf 0}$ origin ($l^{-1} > k_0$), so that such solution involves all directions of the wavevectors in the $(k_x,k_z)$ plane and cannot be considered as paraxial (even when $w \gg k_0^{-1}$). Notably, the wavepacket (\ref{eq19}) is ultrashort (subwavelength) because $k_0 l < 1$, and the $z$-shifts of the centroids look more pronounced in this solution.

Figure~\ref{fig3} shows the real-space evolution of a non-diffracting Bessel wavepacket numerically calculated via Eq.~(\ref{eq18}). 
One can see that the wavepacket indeed propagates with an unvarying elliptical Bessel-like intensity distribution with subluminal velocity. This subluminal velocity allows one to perform a Lorentz $z$-boost to its rest frame, where it becomes a standing circular Bessel wave. Figure~\ref{fig3}(c) shows the noticeable yet tiny fourth-order difference between propagations of the energy and probability centroids.

\section{Conclusions}
We have examined properties of the energy and photon-probability centroids of electromagnetic wavepackets in free space. In the second-order paraxial approximation, these centroids propagate with the same subluminal velocity. This effect is caused by the transverse confinement of the wavepacket and the corresponding tilt of the wavevectors in its Fourier spectrum. A tiny difference between the propagation of the energy and probability centroids appears in the fourth order. Even if the experimental accuracy would allow to measure this difference, it is not clear if there are detectors or light-matter interaction processes specifically sensitive to the photon-probability centroid. We have considered three different types of 2D wavepackets: Gaussian, Bessel, and non-diffracting Bessel. In all cases, the subluminal propagation of their energy centroids is clearly visible in the intensity distributions and can be measured experimentally for propagation distances over several Rayleigh lengths.

Our results shed light on the propagation of electrpmagnetic wavepackets both in the classical-light and single-photon regimes. In the single-photon regime, similar subluminal propagation has been measured in Ref.~\cite{Giovannini2015}, but the continuous-wave regime in this work corresponds to $z$-delocalized monochromatic wave beams where the centroids' positions and propagation are ill-defined. 

It should be noticed that the subluminal propagation of the wavepacket centroids does not contradict  propagation of signals or information transfer with the speed of light. The wavepackets under consideration are solutions infinitely extended in both space and time. When one considers a wavefield which exactly vanishes in the $t<0$ and $z>0$ half-spaces, its non-zero intensity front (precursor) always propagates with the speed of light $c$ \cite{Brillouin_book,Sommerfeld_book}. This is the case even for optical fields in media or waveguides \cite{Brillouin_book,Sommerfeld_book,Pleshko1969,Aaviksoo1991} and for massive quantum waves in free space \cite{Berry2012_II,Bliokh2018}.    


\section*{Acknowledgements}

I am grateful to Prof. Aleksandr Bekshaev for helpful discussions.


\section*{References}

\bibliography{References_STVP}

\end{document}